\documentclass[conference]{IEEEtran}
\usepackage{mathpazo}
\usepackage{times}
\usepackage{color}
\usepackage{amsmath}
\usepackage{amsfonts}
\usepackage{latexsym}
\usepackage{amssymb}
\usepackage{cite}

\usepackage{upref}
\usepackage{theorem}
\usepackage{graphicx}
\usepackage{psfrag}
\usepackage{mathrsfs}
\usepackage{hyperref}

\usepackage{color}

\theoremstyle{plain}
\theorembodyfont{\normalfont\slshape}

\newtheorem{thm}{Theorem$\!$}
\newenvironment{theorem}
{\begin{thm}\hspace*{-1ex}{\bf.}}{\end{thm}}

\newtheorem{lem}[thm]{Lemma$\!$}
\newenvironment{lemma}{\begin{lem}\hspace*{-1ex}{\bf.}}{\end{lem}}

\newtheorem{prop}[thm]{Proposition$\!$}

\newtheorem{cor}[thm]{Corollary$\!$}
\newenvironment{corollary}{\begin{cor}\hspace*{-1ex}{\bf.}}{\end{cor}}

\newtheorem{defn}[thm]{Definition$\!$}

\newtheorem{xmpl}[thm]{Example$\!$}
\newenvironment{example}{\begin{xmpl}\hspace*{-1ex}{\bf.}}{\hfill$\Box$\end{xmpl}}

\newtheorem{cnstr}{Construction$\!$}

\newcounter{enumrom}
\renewcommand{\theenumrom}{(\roman{enumrom})}

\makeatletter
\renewcommand{\@endtheorem}{\endtrivlist}
\makeatother

\makeatletter
\renewcommand{\thefigure}{{\@arabic\c@figure}}
\renewcommand{\fnum@figure}{{\bf Figure\,\thefigure}}
\makeatother

\newcommand{\cG}{\mathcal{G}}

\newcommand{\cT}{\mathcal{T}}

\newcommand{\mathset}[1]{\left\{#1\right\}}
\newcommand{\abs}[1]{\left|#1\right|}

\newcommand{\parenv}[1]{\left( #1 \right)}

\newcommand{\be}[1]{\begin{equation}\label{#1}}
\newcommand{\ee}{\end{equation}}

\renewcommand{\le}{\leqslant}
\renewcommand{\leq}{\leqslant}
\renewcommand{\ge}{\geqslant}
\renewcommand{\geq}{\geqslant}

\newcommand{\Cref}[1]{Co\-ro\-lla\-ry\,\ref{#1}}

\newcommand{\ccap}{\mathsf{cap}}
\newcommand{\send}{S^{\mathrm{end}}}
\newcommand{\sadj}{S^{\mathrm{tan}}}
\newcommand{\tend}{T^{\mathrm{end}}}
\newcommand{\tadj}{T^{\mathrm{tan}}}
\newcommand{\ctend}{\cT^{\mathrm{end}}}
\newcommand{\ctadj}{\cT^{\mathrm{tan}}}
\newcommand{\occ}[2]{n_{#1}\parenv{#2}}
\newcommand{\ma}{T^{\mathrm{rt}}}
\newcommand{\Ma}{\cT^{\mathrm{rt}}}
\newcommand{\mma}{S^{\mathrm{rt}}}
\newcommand{\rg}{T^{\mathrm{gap}}}
\newcommand{\Rg}{\cT^{\mathrm{gap}}}
\newcommand{\rrg}{S^{\mathrm{gap}}}
\newcommand{\rev}[1]{{#1}^{\scriptscriptstyle{R}}}
\newcommand{\comment}[1]{\textcolor{red}{#1}}
\IEEEoverridecommandlockouts
\newcommand{\ignore}[1]{}

\outer\def\proclaim #1. #2\par{\medbreak
 \noindent{\bf#1.\enspace}{\sl#2\par}%
 \ifdim\lastskip<\medskipamount \removelastskip\penalty55\medskip\fi}

\begin{document}


\title{\Huge\bf The Capacity of String-Replication Systems}

\author{
\IEEEauthorblockN{\textbf{Farzad Farnoud (Hassanzadeh)}}
\IEEEauthorblockA{Electrical Engineering  \\
California Institute of Technology \\
Pasadena, CA 91125, U.S.A. \\
{\it farnoud@caltech.edu}}
\and
\IEEEauthorblockN{\textbf{Moshe Schwartz}}
\IEEEauthorblockA{Electrical and Computer Engineering  \\
Ben-Gurion University of the Negev\\
Beer Sheva 8410501, Israel \\
{\it schwartz@ee.bgu.ac.il}}
\and
\IEEEauthorblockN{\textbf{Jehoshua Bruck}}
\IEEEauthorblockA{Electrical Engineering  \\
California Institute of Technology \\
Pasadena, CA 91125, U.S.A. \\
{\it bruck@paradise.caltech.edu}}%
\thanks{This work was supported in part by the NSF Expeditions in Computing Program (The Molecular Programming Project).}
}
\maketitle

\begin{abstract}
It is known that the majority of the human genome consists of repeated
sequences. Furthermore, it is believed that a significant part of the
rest of the genome also originated from repeated sequences and has
mutated to its current form. In this paper, we investigate the
possibility of constructing an exponentially large number of sequences
from a short initial sequence and simple replication rules, including
those resembling genomic replication processes. In other words, our
goal is to find out the capacity, or the expressive power, of these
string-replication systems. Our results include exact capacities, and
bounds on the capacities, of four fundamental string-replication
systems.
\end{abstract}

\section{Introduction}
More than 50\% of the human genome consists of \emph{repeated
  sequences}~\cite{lander2001initial}. An important class of these
repeated sequences are \emph{interspersed repeats}, which are caused
by \emph{transposons}. A transposon, or a ``jumping gene'', is a
segment of DNA that can ``copy and paste'' or ``cut and paste'' itself
into new positions of the genome. Currently, 45\% of the human genome
is known to consist of transposon-driven
repeats~\cite{lander2001initial}.

A second type of repeats are \emph{tandem repeats}, generally thought to be caused by \emph{slipped-strand mispairings}~\cite{mundy2004}. A slipped-strand mispairing is said to occur when, during DNA synthesis, one strand in a DNA duplex becomes misaligned with the other. These mispairings may lead to deletions or insertion of a repeated sequence~\cite{Levinson1987}. While tandem repeats are known to constitute only 3\% of the human genome, they cause important phenomena such as chromosome fragility, expansion diseases, silencing genes~\cite{usdin2008biological}, and rapid morphological variation~\cite{fondon2004}.

While interspersed repeats and random repeats together account for a significant part of the human genome, it is likely that a substantial portion of the unique genome, the part that is not known to contain repeated sequences, also has its origins in ancient repeated sequences that  are no longer recognizable due to change over time~\cite{lander2001initial, usdin2008biological}.

Motivated by the prevalence and the significance of repeated sequences and the fact that much of our unique DNA was likely originally repeated sequences, in this paper we study the \emph{capacity} of \emph{string-replication systems} with simple replication rules including rules that resemble the repeat-producing genomic processes, namely duplication of transposons and duplication caused by slipped-strand mispairings. A string-replication system, to be defined formally later, consists of a set of rewriting rules, an initial sequence, and all sequences that can be obtained by a applying the rules to the initial sequence a finite number of times. The notion of capacity, defined later in the paper,  represents the average number of bits per symbol that can asymptotically be encoded by the sequences in a string-replication system, and thus illustrates the expressive power and the diversity of that system.

In this paper, we consider four replication rules. The first is the \emph{end replication} rule, which allows substrings of a certain length $k$ to be appended to the end of previous sequences. For example, if $k=3$ we may construct the sequence $\mathsf{T\underline{CAT}GC\underline{CAT}}$ from $\mathsf{TCATGC}$. While this rule is not biologically motivated, we present it first because of the simplicity of proving the related results. In particular, we show that nearly all sequences with the same alphabet as the initial sequence can be generated with this rule.

The second rule is called \emph{tandem replication} and allows a substring of length $k$ to be replicated next to its original position. For example, for $k=3$, from the sequence $\mathsf{TCATGC}$, one can generate $\mathsf{T\underline{CAT}\hspace{.2mm} \underline{CAT}GC}$. We show that this rule has capacity zero regardless of the initial sequence. However, if one allows substrings of all length larger than a given value to be copied, the capacity becomes positive except in trivial cases.

The third rule is \emph{reversed tandem replication}, which is similar to tandem replication except that the copy is reversed before insertion. For example, in our previous example, the sequence $\mathsf{T\underline{CAT}\hspace{.2mm} \underline{TAC}GC}$ can be generated. Here, the capacity is zero only in the trivial case in which the initial sequence consists of only one unique symbol.

The last rule is \emph{replication with a gap}, where the copy of a substring of a given length $k$ can be inserted after $k'$ symbols. This rule is motivated by the fact that transposons may insert themselves in places far from their original positions. As an example, for $k=3$ and $k'=1$, from $\mathsf{TCATGC}$, one can obtain $\mathsf{T\underline{CAT}G\underline{CAT}C}$. For this rule, we show that the capacity is zero if and only if the initial sequence is periodic with period equal to the greatest common divisor of $k$ and $k'$.

We note that tandem replication has been already studied in a series
of papers
\cite{dassow1999regularity,dassow2002operations,leupold2004formal,leupold2005uniformly}. However,
this was done in the context of the theory of formal languages, and
the goal of these studies was mainly to determine their place in the
Chomsky hierarchy of formal languages.

In the next section, we present the preliminaries and in the following four sections, we present the results for each of the aforementioned replication rules.
\section{Preliminaries}


Let $\Sigma$ be some finite alphabet. We recall some useful notation commonly used in the theory of formal languages. An $n$-string $x=x_1 x_2 \dots x_{n}\in\Sigma^{n}$ is a finite sequence of alphabet symbols, $x_{i}\in\Sigma$. We say $n$ is the length of $x$ and denote it by $\abs{x}=n$. For two strings, $x\in\Sigma^{n}$ and $y\in\Sigma^{m}$, their concatenation is denoted by $xy\in\Sigma^{n+m}$. The set of all finite strings over the alphabet $\Sigma$ is denoted by $\Sigma^*$. We say $v\in\Sigma^*$ is a \emph{substring} of $x$ if $x=uvw$, where $u,w\in\Sigma^*$. The \emph{alpha-representation} of a string $s$, denoted by $R(s)$, is the set of all letters from $\Sigma$ making up $s$. Thus, $R(s)\subseteq\Sigma$.  The \emph{alpha-diversity} of $s$ is the size of the alpha-representation of $s$, denoted by $\delta(s)=\abs{R(s)}$. Furthermore, let the number of occurrences of a symbol $a\in\Sigma$ in a sequence $s\in\Sigma^*$ be denoted by $n_{x}(a)$. The unique empty word of length $0$ is denoted by $\epsilon$. 

Given a set $S\subseteq \Sigma^*$, we denote
\[S^* = \mathset{ w_1w_2\dots w_m ~|~ w_i\in S, m \geq 0},\]
whereas
\[S^+ = \mathset{ w_1w_2\dots w_m ~|~ w_i\in S, m \geq 1}.\]
For any $x\in \Sigma^*$, $\abs{x}=n\geq m$, the $m$-suffix of $x$ is $w\in \Sigma^m$, such that $x=vw$ for some $v\in\Sigma^*$. Similarly, the $m$-prefix of $x$ is $u\in\Sigma^m$, where $x=uv$ for some
$u\in\Sigma^*$.

A \emph{string system} $S$ is a subset $S\subseteq\Sigma^*$.  For
any integer $n$, we denote by $N_{S}(n)$ the set of length $n$ strings
in $S$, i.e.,
\[
N_{S}(n)=\abs{S\cap\Sigma^{n}}.
\]

The \emph{capacity} of a string system $S$ is defined by 
\[
\ccap(S)=\limsup_{n\to\infty}\frac{\log_{2}N_{S}(n)}{n}.
\]

A \emph{string-replication system} is a tuple $S=(\Sigma,s,\mathcal{T})$, where $\Sigma$ is a finite alphabet, $s\in\Sigma^*$ is a finite string (which we will use to start the replication process), and where $\mathcal{T}$ is a set of functions such that each $T\in\mathcal{T}$ is a mapping from $\Sigma^*$ to $\Sigma^*$ that defines a string-replication rule. The resulting string system $S$, induced by $(\Sigma,s,\mathcal{T})$, is defined as the closure of the string-replication functions $\mathcal{T}$ on the initial string set $\mathset{s}$, i.e., $S$ is the minimal set for which $s\in S$, and for each $s'\in S$ and $T\in\mathcal{T}$ we also have $T(s')\in S$.



\section{End Replication}

We define the end-replication function, $\tend_{i,k}:\Sigma^*\to\Sigma^*$,
as follows: 
\[
\tend_{i,k}(x)=\begin{cases}
uvwv & \text{if \ensuremath{x=uvw}, \ensuremath{\abs{u}=i}, \ensuremath{\abs{v}=k}}\\
x & \text{otherwise.}
\end{cases}
\]
We also define two sets of these functions which will be used later:
\begin{align*}
\ctend_{k} & =\mathset{\tend_{i,k}~|~i\geq0}\\
\ctend_{\geq k} & =\mathset{\tend_{i,k'}~|~i\geq0,k'\geq k}
\end{align*}

Intuitively, in the end-replication system, the transformations
replicate a substring of length $k$ and append the replicated
substring to the end of the original string.
\begin{theorem}
Let $\Sigma$ be any finite alphabet, $k\geq1$ any integer, and $s\in\Sigma^*$,
$\abs{s}\geq k$. Then for $\send_{k}=(\Sigma,s,\ctend_{k})$, 
\[
\ccap(\send_{k})=\log_{2}\delta(s).
\]
 \end{theorem}
\begin{IEEEproof}
First we note that by requiring $\abs{s}\geq k$ we avoid the degenerate
case of $\send_{k}$ containing only $s$. We further note that, by
the definition of the replication functions, 
\[
R(x)=R(\tend_{i,k}(x))
\]
for all non-negative integers $i$ and $k$, and thus, all the strings
in $\send_{k}$ have the same alpha-representation. Thus, trivially,
\[
\ccap(\send_{k})\leq\log_{2}\delta(s).
\]

We now turn to prove the inequality in the other direction. We contend that given a string $x\in\Sigma^*$, $\abs{x}\geq k$, and some string $w\in\Sigma^{k}$, $R(w)\subseteq R(x)$,  with at most $2k$ replication steps we can obtain from $x$ a string $y\in\Sigma^*$ ending with $w$, i.e., $y=vw$.

As a first step, we replicate the prefix of $x$, i.e., if $x=uv$,
$\abs{u}=k$, then 
\[
x'=\tend_{0,k}(x)=uvu.
\]
By doing so we ensure that for any symbol $a\in R(x)$ there is a
$k$-substring of $x'$ starting with $a$, and a $k$-substring of
$x'$ ending with $a$.

Let us now denote the symbols of $w$ as $w=w_{1}w_{2}\dots w_{k}$, $w_i\in\Sigma$.  Assume that the $k$-substring of $x'$ starting at position $i_{1}$ ends with $w_{1}$. We form
\[
x_1 =\tend_{i_{1}-1,k}(x')
\]
whose $1$-suffix is just $w_{1}$. Next, assume the $k$-substring of $x'$ starting at position $i_{2}$ starts with $w_{2}$. Note that $x'$ is a prefix of $x_1$. We form
\[
x_2 =\tend_{\abs{x_1}-k+1,k}\parenv{\tend_{i_{2}-1,k}(x_1 )}.
\]
It easy to verify $x_2 $ has a $2$-suffix of $w_{1}w_{2}$. Continuing in the same way, let $i_j $ be starting position of a $k$-substring of $x'$ starting with $w_j $. We form 
\[
x_j =\tend_{\abs{x_{j-1}}-k+1,k}\parenv{\tend_{i_j -1,k}(x_{j-1})},
\]
for $j=3,\dots,k$. Note that $x_j$ has a $j$-suffix $w_1,\dotsc,w_j$.

It follows that after $2k$ replication steps we can obtain from any such
$x$ a string with any given $k$-suffix $w$, provided $R(w)\subseteq R(x)$.
Thus, from the initial string $s$, we can obtain a string $s'$ with
all of the strings of $R(s)^{k}$ appearing as $k$-substrings, using
at most $2k\delta(s)^{k}$ replication steps%
\footnote{This bound may be improved, but this will not affect the capacity
calculation.%
}, i.e., 
\[
\abs{s'}\leq\abs{s}+2k^{2}\delta(s)^{k}.
\]

After having obtained $s'$, each replication may replicate any of the
$k$-strings in $R(s)^{k}$ in a single operation. Thus, for all
$n=\abs{s'}+tk$, $t$ a non-negative integer, the number of distinct
strings in $\send_{k}$ is bounded from below by
\[
N_{\send_{k}}(n)\geq\delta(s)^{n-\abs{s'}}.
\]
Since $\abs{s'}$ is a constant, we have 
\[
\ccap(\send_{k})\geq\log_{2}\delta(s).
\]

\end{IEEEproof}
The following is an obvious corollary. 
\begin{theorem}
Let $\Sigma$ be any finite alphabet, $k\geq1$ any integer, and $s\in\Sigma^*$,
$\abs{s}\geq k$. Then for $\send_{\geq k}=(\Sigma,s,\ctend_{\geq k})$,
\[
\ccap(\send_{\geq k})=\ccap(\send_{k})=\log_{2}\delta(s).
\]
\end{theorem}
\begin{IEEEproof}
Since for all $n\geq k$,
\[ N_{\send_{k}}(n)\leq N_{\send_{\geq k}}(n) \leq \delta(s)^n,\]
the claim follows.
\end{IEEEproof}

\section{Tandem Replication}

We now consider different replication rules, $\tadj_{i,k}:\Sigma^*\to\Sigma^*$,
defined by 
\[
\tadj_{i,k}(x)=\begin{cases}
uvvw & \text{if \ensuremath{x=uvw}, \ensuremath{\abs{u}=i}, \ensuremath{\abs{v}=k}}\\
x & \text{otherwise.}
\end{cases}
\]
We also define the sets 
\begin{align*}
\ctadj_{k} & =\mathset{\tadj_{i,k}~|~i\geq0}\\
\ctadj_{\geq k} & =\mathset{\tadj_{i,k'}~|~i\geq0,k'\geq k}
\end{align*}

Unlike the end replication discussed in the previous section, tandem
replication takes a $k$-substring and replicates it adjacent to itself
in the string. Also, the capacity of tandem-replication systems
is in complete contrast to end-replication systems.
\begin{theorem}
Let $\Sigma$ be any finite alphabet, $k$ any positive integer, and $s\in\Sigma^*$, with
$\abs{s}\geq k$. Then for $\sadj_{k}=(\Sigma,s,\ctadj_{k})$, 
\[
\ccap(\sadj_{k})=0.
\]
 \end{theorem}
\begin{IEEEproof}
Consider any $n$-string $x\in\Sigma^*$, $\abs{x}\geq k$. Instead
of viewing $x=x_1 x_2 \dots x_{n}$ as a sequence of $n$ symbols
from $\Sigma$, we can, by abuse of notation, view it as a sequence
of $n-k+1$ overlapping $k$-substrings $x=x'_{1}x'_{2}\dots x'_{n-k+1}$,
where 
\[
x'_{i}=x_{i}x_{i+1}\dots x_{i+k-1}.
\]

For a $k$-string $y=y_{1}y_{2}\dots y_{k}$, $y_i\in\Sigma$, its cyclic shift by
one position is denoted by 
\[
Ey=y_{2}y_{3}\dots y_{k}y_{1}.
\]
A cyclic shift by $j$ positions is denoted by 
\[
E^{j}y=y_{j+1}y_{j+2}\dots y_{k}y_{1}y_{2}\dots y_j .
\]
We say two $k$-strings, $y,z\in\Sigma^{k}$, are cyclically equivalent
if 
\[
y=E^{j}z,
\]
for some integer $j$. Clearly this is an equivalence relation. Let
$\phi(y)$ denote the equivalence class of $y$. If $y$ and $z$
are cyclically equivalent, then $\phi(y)=\phi(z)$.

We now define 
\[
\Phi(x)=\phi(x'_{1})\phi(x'_{2})\dots\phi(x'_{n-k+1}),
\]
i.e., $\Phi(x)$ is the image of the overlapping $k$-substrings of
$x$ under $\phi$. We also observe that knowing $x'_{1}$ and $\Phi(x)$
enables a full reconstruction of $x$.

At this point we turn to consider the effect of the replication
$\tadj_{i,k}$ on a string $x\in\Sigma^*$, $\abs{x}\geq k$. When viewed
as a sequence of overlapping $k$-substrings, as defined above,
\[
\tadj_{i,k}(x)=x'_{1}\dots x'_{i-1}x'_{i}Ex'_{i}E^{2}x'_{i}\dots E^{k-1}x'_{i}x'_{i}x'_{i+1}\dots x'_{n-k+1}.
\]
Since $\phi(x'_{i})=\phi(E^{j}(x'_{i}))$ for all $j$, we have 
\begin{align*}
\Phi(\tadj_{i,k}(x)) & =\phi(x'_{1})\dots\phi(x'_{i-1})\\
 & \quad\ \phi(x'_{i})\phi(x'_{i})\dots\phi(x'_{i})\\
 & \quad\ \phi(x'_{i+1})\dots\phi(x'_{n-k+1}),
\end{align*}
where $\phi(x'_{i})$ appears $k+1$ consecutive times.

Thus, we may think of $\phi(x'_{i})$ as a bin, and the action of
$\tadj_{i,k}$ as throwing $k$ balls into the bin $\phi(x'_{i})$.
The number of bins does not change throughout the process, and is
equal to one more than the number of times $\phi(x'_{i})\neq\phi(x'_{i+1})$,
where $x=s$ is the original string. If $b$ is the number of bins
defined by $s$, then the number of strings obtained by $m$ replications
is exactly $\binom{b+m-1}{b-1}$. Since this number grows only polynomially
in the length of the resulting string, we have 
\[
\ccap(\sadj_{k})=0.
\]

\end{IEEEproof}
When considering $\sadj_{\geq k}=(\Sigma,s,\tadj_{\geq k})$ the situation
appears to be harder to analyze.
\begin{theorem}
\label{th:lb1}
For any finite alphabet $\Sigma$, and any string $s\in\Sigma^*$ of nontrivial alpha-diversity, $\delta(s)\geq2$,
we have 
\[
\ccap(\sadj_{\geq1})\geq\log_{2}(r+1),
\]
where $r$ is the largest (real) root of the polynomial 
\[
f(x)=x^{\delta(s)}-\sum_{i=0}^{\delta(s)-2}x^{i}.
\]
\end{theorem}
\begin{IEEEproof}
The proof strategy is the following: we shall show that $\sadj_{\geq
  1}$ contains, among other things, a regular language. The capacity
of that regular language will serve as the lower bound we claim.

For the first phase of the proof, assume $i_{1}<i_{2}<\dots<i_{\delta(s)}$
are the indices of $\delta(s)$ distinct alphabet symbols in $s$.
We produce a sequence of strings, $s_{0}=s,s_{1},\dots,s_{\delta(s)-1}$,
defined iteratively by 
\[
s_j =\tadj_{i_{\delta(s)-j}-1,i_{\delta(s)}-i_{\delta(s)-j}+j}(s_{j-1}),
\]
for $j=1,2,\dots,\delta(s)-1$. After this set of steps, the
$\delta(s)$-substring starting at position $i_{\delta(s)}$ of
$s_{\delta(s)-1}$ contains $\delta(s)$ distinct symbols. In what
follows we will only use these symbols for replication, and thus, the
constant amount of other symbols in $s_{\delta(s)-1}$ does not affect
the capacity calculation. Thus, for ease of presentation we shall
assume from now on that $\abs{s}=\delta(s)$, i.e., the initial string
contains no repeated symbol from the alphabet. Furthermore, without
loss of generality, let us assume these symbols are
$a_{\delta(s)},a_{\delta(s)-1},\dots,a_{1}$, in this order.

We now perform the following iterations: In iteration $i$, where
$i=\delta(s),\delta(s)-1,\dots,2$, we replicate $i$-substrings equal
to $a_ia_{i-1}\dots a_2a_1$. As a final iteration, we may replicate $1$-substrings
without constraining their content. It is easy to verify the resulting
strings form the following regular language,
\[
S=\parenv{a_{\delta(s)}^{+}\parenv{a_{\delta(s)-1}^{+}\parenv{\dots\parenv{a_2^{+}\parenv{a_1^{+}}^{+}}^{+}}^{+}}^{+}}^{+}.
\]
The construction process implies $S \subseteq \sadj_{\geq 1}$.
\begin{figure}[ht]
\psfrag{aaa}{$a_1$}
\psfrag{bbb}{$a_2$}
\psfrag{ccc}{$a_3$}
\psfrag{ddd}{$a_{\delta(s)-1}$}
\psfrag{eee}{$a_{\delta(s)}$}
\centering
\includegraphics[scale=0.5]{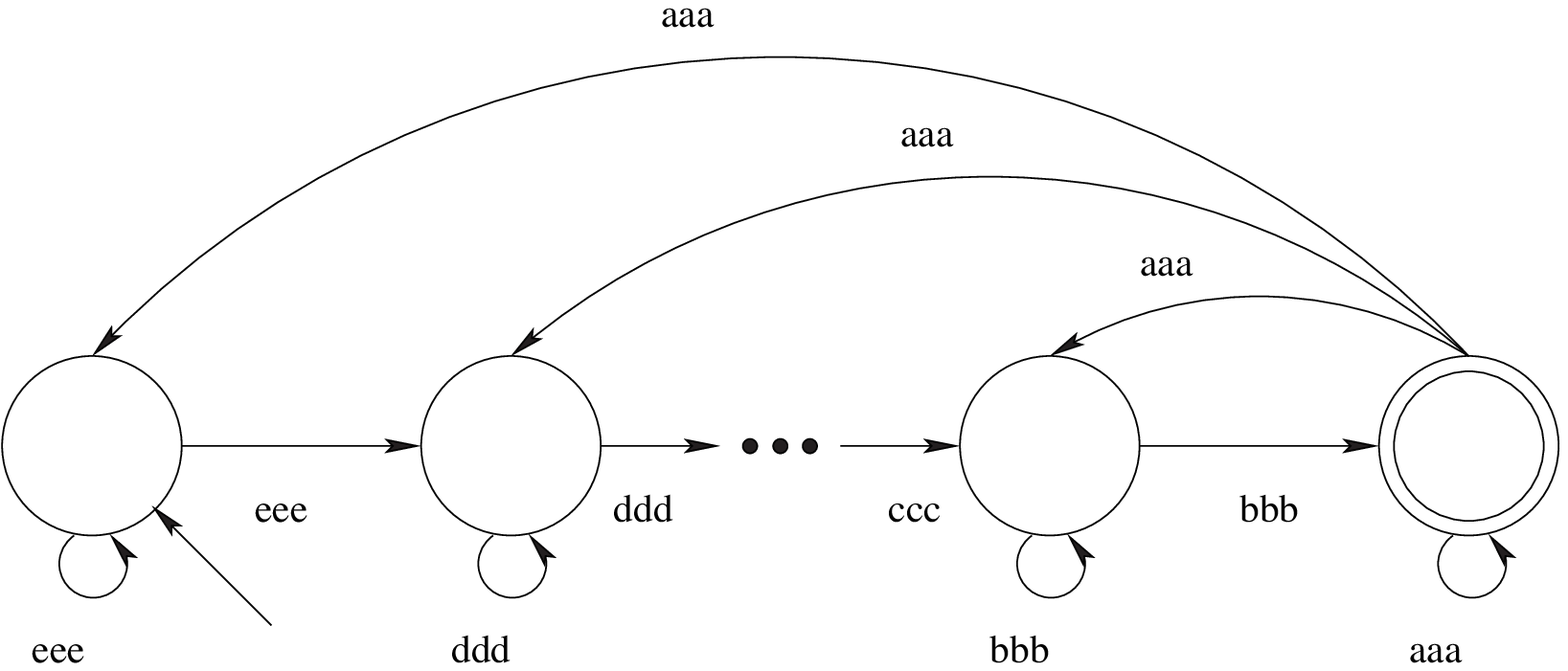}
\caption{
  The finite-state automaton accepting the regular language used in the
  proof of Theorem \ref{th:lb1}.
}
\label{fig:adjgraph}
\end{figure}
\ignore{
Furthermore, without loss of generality, let us assume these symbols are $\mathset{1,2,\dots,\delta(s)}$, sorted in descending order.

We now perform the following iterations: In iteration $i$, where
$i=\delta(s),\delta(s)-1,\dots,2$, we replicate $i$-substrings equal
to $i,i-1,\dots,2,1$. As a final iteration, we may replicate $1$-substrings
without constraining their content. It is easy to verify the resulting
strings form the following regular language,
\[
S=\parenv{\delta(s)^{+}\parenv{\comment{(}\delta(s)-1\comment{)}^{+}\parenv{\dots\parenv{2^{+}\parenv{1^{+}}^{+}}^{+}}^{+}}^{+}}^{+}.
\]
The construction process implies $S \subseteq \sadj_{\geq 1}$.

\begin{figure}[ht]
\psfrag{aaa}{$1$}
\psfrag{bbb}{$2$}
\psfrag{ccc}{$3$}
\psfrag{ddd}{$\delta(s)-1$}
\psfrag{eee}{$\delta(s)$}
\centering
\includegraphics[scale=0.5]{adj_graph.eps}
\caption{
  The finite-state automaton accepting the regular language used in the
  proof of Theorem \ref{th:lb1}.
}
\label{fig:adjgraph}
\end{figure}
}

The finite-state automaton accepting $S$ is depicted in Figure~\ref{fig:adjgraph}. The graph is primitive and lossless, and thus, for the purpose of calculating the capacity, instead of counting the number of length $n$ words in $S$, we can count the number of length $n$ paths in the automaton graph $\cG$ (see \cite{LinMar85,Imm04}). By Perron-Frobenius theory,
\[\ccap(\sadj_{\geq 1})\geq \ccap(S) = \log_2 \lambda(A_\cG),\]
where $\lambda(A_\cG)$ is the largest magnitude of an eigenvalue of
$A_\cG$, and where $A_{\cG}$ denotes the adjacency matrix of $\cG$. We
note that $A_\cG$ is the $\delta(s)\times\delta(s)$ matrix
\[
A_\cG=\begin{pmatrix}1 & 1\\
 & 1 & 1\\
 &  & 1 & 1\\
 &  &  & \ddots & \ddots\\
 &  &  &  & 1 & 1\\
1 & 1 & 1 & \dots & 1 & 1
\end{pmatrix},
\]
and its largest eigenvalue is the largest real root of 
\[
\det(\lambda I-A_\cG)=(\lambda-1)^{\delta(s)}-\sum_{i=0}^{\delta(s)-2}(\lambda-1)^{i}.
\]
Setting $x=\lambda-1$ we obtain the desired result.
\end{IEEEproof}

At least in one case, the bound of Theorem \ref{th:lb1} is attained with
equality, as is shown in the following corollary.
\begin{corollary}
For $\Sigma=\mathset{0,1}$, and $s\in\Sigma^*$ with $\delta(s)=2$
we have 
\[
\ccap(\sadj_{\geq1})=1.
\]
\end{corollary}
\begin{IEEEproof}
By applying Theorem \ref{th:lb1} we get 
\[
\ccap(\sadj_{\geq1})\geq1.
\]
We also have the trivial upper bound 
\[
\ccap(\sadj_{\geq1})\leq\log_{2}\abs{\Sigma}=1,
\]
which completes the proof. \end{IEEEproof}

For $\sadj_{\geq k}$ and general $k$, we claim a weaker result, that
is provided in the following theorem.
\begin{theorem}
For any finite alphabet $\Sigma$, and any binary string
$s\in\Sigma^*$, $\abs{s}\geq k$, of nontrivial alpha-diversity,
$\delta(s)\geq2$, we have
\[
\ccap(\sadj_{\geq k})\geq\log_{2}r>0,
\]
where $r$ is the largest root of the polynomial 
\[
f(x)=x^{k+1}-x-1.
\]
\end{theorem}
\begin{IEEEproof}
The proof strategy is, again, to find a regular language that is a
subset of $\sadj_{\geq k}$ and use its capacity as a lower bound.
We start with the following preparation, by performing the following $k$
replications,
\[ s'=\tadj_{0,2k-1}\parenv{\dots\parenv{\tadj_{0,k+1}\parenv{\tadj_{0,k}(s)}}}.\]
If we denote $s'=s'_1 s'_2 \dots$, where $s'_i\in \Sigma$, then it is easy to verify
that
\[s'_{k+1}=s'_{k+2}=\dots=s'_{2k}=s'_1.\]
Since $\delta(s')=\delta(s)\geq 2$, this run of at least $k$ consecutive equal symbols, must end. Without loss of generality, assume $0,1\in \Sigma$, and (possibly after an appropriate relabeling of the symbol names) either $0^k 1$ or $1 0^k$ form a substring of $s'$. We shall assume the former, and the proof for the latter case is similar. We ignore the rest of the symbols, as they will not affect the capacity. Thus, we may proceed as if the initial string $s$ is $0^k 1$.

We now generate more strings by replicating only substrings of the
form $0^{k}1$ or $0^{k-1}1$. The resulting set of strings contains the
regular language
\[
S=\parenv{\parenv{0^{k}1}^{+}\parenv{0^{k-1}1}^{+}}^{+}.
\]
We can follow the same steps as in the proof of Theorem \ref{th:lb1}
in order to find the capacity of $S$.  It is given by the base-$2$
logarithm of the largest real solution for the equation
\[
x^{-(k+1)}+x^{-k}=1.
\]
By rearranging, we get the claim. It is also easy to verify that the
claimed $r$ satisfies $r>1$, and so the capacity is strictly positive. 
\end{IEEEproof}

\section{Reversed Tandem Replication}

Consider the reversed tandem replication rule $\ma_{i,k}:\Sigma^*\to\Sigma^*$
defined as
\[
\ma_{i,k}(x)=\begin{cases}
uv\rev{v}w & \text{if $x=uvw$, $\abs{u}=i$, $\abs{v}=k$,}\\
x & \text{otherwise},
\end{cases}
\]
where $\rev{y}$ is the reverse of $y$, i.e.,
$\rev{y}=y_{m}y_{m-1}\dots y_{1}$ for a sequence $y=y_{1}y_{2}\dots
y_{m}\in\Sigma^*$.  Furthermore, let
\[
\Ma_{k}=\mathset{\left. \ma_{i,k} ~\right|~ i\ge 0 }.
\]
and use $\mma_{k}=(\Sigma,s,\Ma_k)$. Since the starting string $s$ will play
a crucial role, we shall often use the notation $\mma_k(s)$.
\begin{lemma}
\label{lem:mirror1}
Let $s\in \Sigma^k$ such that $s\neq\rev{s}$. Then
\[
\ccap(\mma_{k}(s))\ge\frac{1}{k}.
\]
\end{lemma}
\begin{IEEEproof}
By repeatedly applying replication to the last block of $k$ symbols, we can create any sequence of alternating blocks $s$ and $\rev{s}$, starting with $s$. To extend any run of $s$ (resp.\ $\rev{s}$), except the first one, we can apply replication to the last block of the previous run, which is an $\rev{s}$ block (resp.\ $s$). Thus, the regular language
\[S=s \rev{s} \mathset{s,\rev{s}}^*,\]
satisfies $S\subseteq \mma_k(s)$.
Since $s\neq \rev{s}$, we easily see that
\[\ccap(\mma_k(s))\geq \ccap(S) = \frac{1}{k}.\]
\end{IEEEproof}
Note that the requirement that $s\neq\rev{s}$ implies that $k\ge2$.

The following theorem states that the capacity of reversed tandem replication is positive except in trivial cases.
\begin{theorem}
\label{thm:mirror2}
For any $s\in \Sigma^*$, $\abs{s}\geq k$, we have
$\ccap(\mma_{k}(s))=0$ if and only if $\delta(s)=1$.
\end{theorem}
\begin{IEEEproof}
It is clear that if $\delta\parenv{s}=1$, then $\ccap(\mma_{k}(s))=0$.
For the other direction, suppose that $\ccap(\mma_{k}(s))=0$. We show
that $\delta(s)=1$. We first prove this for $\abs{s}=k$.

Denote $s=s_1s_2\dots s_k$, with $s_i\in \Sigma$.
Since $\ccap(\mma_{k}(s))=0$, by Lemma \ref{lem:mirror1}, we
have that $s=\rev{s}$, or equivalently,
\begin{equation}
s_{i}=s_{k+1-i},\quad \forall i\in[k].\label{eq:mirror1}
\end{equation}
From $\ccap(\mma_{k}(s))=0$, it also follows that
$\ccap(\mma_{k}(s\rev{s}))=0$, which in turn implies that
$\ccap(\mma_{k}(s_{2}s_{3}\dots s_{k}s_{k}))=0$.  Hence,
\begin{align*}
s_{2} & =s_{k},\\
s_{i+1} & =s_{k+2-i},\quad \forall 2\le i\le k-1.
\end{align*}
or equivalently, 
\begin{align}
s_{2} & =s_{k},\label{eq:mirror2}\\
s_{i+2} & =s_{k+1-i},\quad \forall i\in[k-2].\label{eq:mirror3}
\end{align}
From (\ref{eq:mirror1}) and (\ref{eq:mirror3}), it follows that
\begin{equation}
s_{i}=s_{i+2},\quad \forall i\in[k-2].\label{eq:mirror4}
\end{equation}
It is also true that $s_{1}=s_{2}$ since $s_{1}=s_{k}$ from (\ref{eq:mirror1})
and $s_{2}=s_{k}$ from (\ref{eq:mirror2}). The expressions (\ref{eq:mirror4})
and $s_{1}=s_{2}$ prove that $\delta(s)=1$.

Finally, let $s\in \Sigma^*$ be such that $\abs{s}\geq k$. If $s'$ is
a $k$-substring of $s$, then obviously
\[\ccap(\mma_k(s))\geq \ccap(\mma_k(s')).\]
Since we have $\ccap(\mma_k(s))=0$, then $\ccap(\mma_k(s'))=0$, and
using the above proof for length $k$ strings, we get
$\delta(s')=1$. Since this is true for every $k$-substring $s'$ of $s$, we
must have $\delta(s)=1$.
\end{IEEEproof}

In Theorem~\ref{thm:mirror3}, we show that in determining the capacity
of a system $\mma_k(s)$, only $\delta(s)$ is important and not the
actual sequence $s$. The idea behind the proof is that any other
finite sequence with alphabet $R(s)$ appears as a substring of some
sequence in $\mma_{k}(s)$.  To show this, we use the following lemma
in the proof of Theorem~\ref{thm:mirror3}.
\begin{lemma}
\label{lem:mirror2.5}
For any $x,y\in \Sigma^*$, with $\abs{y}\ge k$,
if for all $a\in\Sigma$, $\occ ya\ge\occ xa$, then $x$ is a suffix
of some sequence in $\mma_{k}(y)$.
\end{lemma}
\begin{IEEEproof}
Since we can increase the length of $y$ by applying the function
$\ma_{0,k}$, while maintaining $\occ ya\ge\occ xa$ for all $a\in
\Sigma$, we assume without loss of generality that $\abs{y}\ge2k$. We
also assume $\abs{x}>0$, or else the claim is trivial. 

Suppose that the \emph{last} symbol of $x$ is $a$. We construct a sequence
$y''$ from $y$ using the functions $\Ma_{k}$ such that $a$ is the last
symbol of $y''$, i.e., $a$ is ``pushed'' to the end.  Let $i$ be such
that $y_{i}=a$. Consider the conditions
\[i \ge k, \qquad\qquad \abs{y}-i  \ge k. \]
At most one of the two conditions does not hold. If the former does
not hold, let $y'=\ma_{0,k}(y)$. There is a copy of $a$
at position $i'=2k-i+1$ in $y'$, i.e., $y'_{i'}=a$. We
have $i'\ge k$ and $\abs{y'}-i'\ge3k-\parenv{2k-i+1}\ge k$.
If the latter does not hold, let $y'=\ma_{i-k,k}(y)$ and
$i'=i$. If both conditions hold, let $y'=y$ and $i'=i$. We thus
have $y'_{i'}=a$ with $i'\ge k$ and $\abs{y'}-i'\ge k$.
The significance of these conditions is that they enable us to replicate
blocks of length $k$ containing $a$ without the need to concern
ourselves with the boundaries of the sequence.

Let $\abs{y'}-i'=q(k-1)+r$ such that $q$ and $r$
are integers with $q\ge1$ and $0\le r<k-1$.

First, suppose $k$ is even. We let $y''=\ma_{i'-k/2,k}(y')$.
Now there is a copy of $a$ in $y''$ at position $i''=i'+k+1$. The
distance of this copy from the end of $y''$ is 
\[
\abs{y''}-i''=\abs{y'}+k-\parenv{i'+k+1}=\abs{y'}-i'-1.
\]
Hence, the distance is decreased by one, compared with $y'$. We repeat
the same procedure and update $y''$ and $i''$ as
\begin{align*}
y'' & \leftarrow\ma_{i''-k/2,k}\parenv{y''},\\
i'' & \leftarrow i''+k+1,
\end{align*}
 until we have $\abs{y''}-i''=q(k-1)$. At this
point we switch to repeating
\begin{align}
\label{eq:rep1}
y'' & \leftarrow\ma_{i''-1,k}\parenv{y''},\\
\label{eq:rep2}
i'' & \leftarrow i''+2k-1,
\end{align}
until $a$ becomes the last symbol of $y''$. 

Next, suppose that $k$ is odd and $r$ is even. We let
$y''=\ma_{i'-\parenv{k-1}/2}(y')$.  Now there is a copy of $a$ in
$y''$ at position $i''=i'+k+2$. The distance of this copy from the end
of $y''$ is
\[
\abs{y''}-i''=\abs{y'}+k-\parenv{i'+k+2}=\abs{y'}-i'-2.
\]
The distance is thus decreased by two, compared with $y'$. Since $r$
is even, by repeating the same procedure and updating $y''$ and $i''$,
we can have $\abs{y''}-i''=q\parenv{k-1}$.  We then repeat
\eqref{eq:rep1} and \eqref{eq:rep2} until $a$ becomes the last symbol
of $y''$.

Finally, suppose that $k$ and $r$ are both odd. Let
$y''=\ma_{i-1,k}(y')$.  There is a copy of $a$ in $y''$ at position
$i''=i'$. The distance of this copy from the end of $y''$ is
$\abs{y''}-i''=\abs{y'}+k-i$.  Let $\abs{y''}-i''=q'\parenv{k-1}+r'$
where $q'$ and $r'$ are integers with $q'\ge1$ and $0\le r'<k-1$. We
thus have
\[
r'=r+1+\parenv{q+1-q'}\parenv{k-1}.
\]
Since $k-1$ is even and $r$ is odd, we find that $r'$ is even.
We can then proceed as in the previous case in which $k$ is odd and
$r$ is even.

We have shown that any symbol present in $y$ can be ``pushed'' to
the end position. We repeatedly apply the same argument by disregarding
the last element of $y''$ and pushing the next appropriate element
to the end position. The final result is a sequence in $\mma_{k}(y)$
which ends with $x$. \end{IEEEproof}
\begin{theorem}
\label{thm:mirror3}
For all $s\in \Sigma^*$, $\abs{s}\geq k$, $\ccap(\mma_{k}(s))$ depends
on $s$ only through $\delta(s)$.\end{theorem}
\begin{IEEEproof}
Consider two sequences $s,t\in\Sigma^*$, $\abs{s},\abs{t}\geq k$, such
that $\delta(s)=\delta(t)$.  Since the identity of the symbols is
irrelevant to the capacity, we may assume that $R(s)=R(t)$. By
appropriate replications, it is easy to find a sequence
$t'\in\mma_{k}(t)$ such that for all $a\in \Sigma$, we have
$\occ{t'}a\ge\occ sa$.  We then apply Lemma \ref{lem:mirror2.5} and
show that $s$ is a substring of some sequence
$t''\in\mma_{k}(t)$. Hence,
\[\ccap(\mma_{k}(s))\le\ccap(\mma_{k}(t''))\leq \ccap(\mma_{k}(t)).\]
Similarly, we can show that
$\ccap(\mma_{k}(t))\le\ccap(\mma_{k}(s))$.  Hence,
$\ccap(\mma_{k}(s))=\ccap(\mma_{k}(t))$.
\end{IEEEproof}
\begin{example}
Suppose $s$ is a string of length $k$ such that $s=\rev{s}$. We show that for positive integers $p$ and $q$, we have 
\begin{equation}\label{eq:numerical1}
N_{\mma_{k}(s)}(pqk)\ge \left(N_{\mma_{k}(s)}(pk)\right)^q.
\end{equation}
To see this, note that to generate sequences of length $pqk$, we can first generate a sequence of length $qk$ consisting of $q$ copies of $s$, and then from each of these copies, generate a sequence of length $pk$. It is clear that \eqref{eq:numerical1} also holds for the case in which $s$ is equal to a relabeling of $\rev s$, where the relabeling map is bijective, e.g., $s=012$. If we let $q\to \infty$ in \eqref{eq:numerical1}, we find that 
\begin{equation}\label{eq:numerical2}
\ccap\parenv{\mma_{k}(s)} \ge \frac{\log_2 N_{\mma_{k}(s)}(pk)}{pk}\cdot
\end{equation}

Using a computer, we obtain Table \ref{tab:mirr-adj} for the given values of $s$ and $k$, and then use \eqref{eq:numerical2} to find the following lower bounds on the capacity,
\begin{align*}
\ccap\parenv{\mma_{2}(01)} & 
\ge\frac{\log_2 584}{14}\ge0.65,\\
\ccap\parenv{\mma_{3}(010)} &  
\ge\frac{\log_2 2894} {21}\ge0.54,\\
\ccap\parenv{\mma_{3}(012)} &
\ge\frac{\log_2 11577}{21}\ge0.64.
\end{align*}
\end{example}

\begin{table}
\caption{Numerical results for reversed tandem replication\label{tab:mirr-adj}}
\begin{centering}
\begin{tabular}{|c|c|c|c|c|c|c|c|c|}
\hline 
$s=01,k=2$ & $n$ & 2 & 4 & 6 & 8 & 10 & 12 & 14\\
\hline 
 & $N(n)$ & 1 & 1 & 3 & 10 & 37 & 145 & 584\\
\hline 
\hline 
$s=010,k=3$ & $n$ & 3 & 6 & 9 & 12 & 15 & 18 & 21\\
\hline 
 & $N(n)$         & 1 & 1 & 3 & 14 & 78 & 467 & 2894\\
\hline 
\hline 
$s=012,k=3$ & $n$ & 3 & 6 & 9 & 12 & 15 & 18 & 21\\
\hline 
 & $N(n)$ & 1 & 1 & 4 & 25 & 182 & 1423 & 11577\\
\hline 
\end{tabular}
\end{centering}
\end{table}

\section{Replication with a Gap}

Consider the replication-with-a-gap rule $\rg_{i,k,k'}:\Sigma^*\to\Sigma^*$ defined as
\[
\rg_{i,k,k'}(x)=\begin{cases}
uvwvz, & \text{if $x=uvwz$, $\abs{u}=i$,}\\
& \text{$\abs{v}=k$, $\abs{w}=k'$,}\\
x, & \text{otherwise.}
\end{cases}
\]
 Furthermore, we let 
\[
\Rg_{k,k'}=\mathset{\left. \rg_{i,k,k'} ~\right|~ i\ge0},
\]
and use $\rrg_{k,k'}=(\Sigma,s,\Rg_{k,k'})$, for some $s\in \Sigma^*$. We may also use $\rrg_{k,k'}(s)$ to denote the aforementioned string system. To avoid trivialities, throughout this section, we assume $k,k'\ge1$.

For a sequence $s=s_1s_2\dots$, with $s_i\in \Sigma$, we conveniently denote the substring starting at position $i$ and of length $k$ as $s_{i,k}=s_is_{i+1}\dots s_{i+k-1}$. Furthermore, for two sequences of equal length, $s,s'\in \Sigma^k$, we denote their Hamming distance as $d_H(s,s')$, which is the number of coordinates in which $s$ and $s'$ disagree.

\begin{lemma}
\label{lem:hamming}
For all $s\in\Sigma^*$ such that $\abs{s}\geq k+k'$,
we have
\[
\ccap(\rrg_{k,k'}(s))\ge\frac{1}{k}\log_2\parenv{1+d_H\parenv{s_{1,k},(s^2)_{k+1,k}}}.
\]
\end{lemma}
\begin{IEEEproof}
The proof considers two cases: either $k\geq k'$, or $k< k'$. We prove
the former. The proof for the latter is similar. It also suffices to
consider only $\abs{s}=k+k'$, since for longer strings we can simply
ignore the extra symbols.

For simplicity of notation, let $s=x_1 \dots x_{k}y_{1}\dots
y_{k'}$, where $x_i,y_i\in \Sigma$. We initially apply $\rg_{0,k,k'}$ to $s$
and obtain
\[s'=\rg_{0,k,k'}(s) = x_1 \dots x_{k}y_{1}\dots
y_{k'}x_1 \dots x_{k}.\]

We then apply $\rg_{i,k,k'}$ to $s'$, for all $0\leq i\leq k$, and get the
following list of results:
\begin{alignat*}{2}
x_1 \dots x_{k} & \ y_{1}\dots y_{k'} & \ x_1 \dots x_{k} & \ x_1 x_2 \dots x_{k'}x_{k'+1}\dots x_{k}\\
x_1 \dots x_{k} & \ y_{1}\dots y_{k'} & x_1 \dots x_{k} & \ y_{1}x_2 \dots x_{k'}x_{k'+1}\dots x_{k}\\
 &  & \vdots\\
x_1 \dots x_{k} & \ y_{1}\dots y_{k'} & x_1 \dots x_{k} & \ y_{1}y_{2}\dots y_{k'}x_{k'+1}\dots x_{k}\\
x_1 \dots x_{k} & \ y_{1}\dots y_{k'} & x_1 \dots x_{k} & \ y_{1}y_{2}\dots y_{k'}\ x_1 \ \dots\ x_{k}\\
 &  & \vdots\\
x_1 \dots x_{k} & \ y_{1}\dots y_{k'} & x_1 \dots x_{k} & \ y_{1}y_{2}\dots y_{k'}x_1 \dots x_{k-k'}
\end{alignat*}
where the five explicitly stated sequences correspond to $i=0,1,k',k'+1,k$.
From these results, it is clear that we have
$1+d_H(s_{1,k},(s^2)_{k+1,k})$ distinct sequences. Since the same
operation can be repeated, i.e., apply $\rg_{i,k,k'}$ to $s'$, for all
$0\leq i\leq k$, to all the distinct results of the previous round,
the number of sequences in $\rrg_{k,k'}$ with length $2k+k'+ik$ is at
least
\[
N_{\rrg_{k,k'}}(2k+k'+ik)\geq \parenv{1+d_H(s_{1,k},(s^2)_{k+1,k})}^{i}.
\]
This completes the proof.
\end{IEEEproof}
With an example, we show that the lower bound of Lemma~\ref{lem:hamming} is sharp.  Choose $s$ as
\[ s=a_{1}\dots a_{k}ba_{2}\dots a_{k},\]
where $b\neq a_1$. Suppose $t\in\rrg_{k,k}(s)$. Each $k$-substring $t_{(i-1)k+1,k}$, for nonnegative integers $i\le |t|/k$, either equals $a_1\dots a_k$ or $ba_2\dots a_k$. Thus for a nonnegative integer $j$ there are no more than $2^j$ sequence of length $jk$ in $\rrg_{k,k}(s)$. Hence, \[\ccap(\rrg_{k,k}(s))\le \lim_{j\to \infty} \frac{\log_2 2^j}{jk}=\frac{1}{k}\] 
which matches the lower bound given in Lemma \ref{lem:hamming}, and so $\ccap(\rrg_{k,k}(s))=\frac1k$.


The next corollary is an immediate result of the previous lemma.
\begin{corollary}
\label{cor:hamming}
Assume $\ccap(\rrg_{k,k'}(s))=0$, where $s\in\Sigma^*$ and $\abs{s}\geq k+k'$.
For any $(k+k')$-substring of $s$, denoted $x_1 \dots x_{k}y_{1}\dots y_{k'}$,
with $x_i,y_i\in\Sigma$, we have
\begin{align*}
x_1 \dots x_{k} & =y_{1}\dots y_{k'}x_1 \dots x_{k-k'}, & \quad\text{if $k>k'$,}\\
x_1 \dots x_{k} & =y_{1}\dots y_{k}, & \quad\text{if $k\le k'$.}
\end{align*}
\end{corollary}

This corollary is used in the following theorem.
\begin{theorem}
For $s\in\Sigma^*$, $\abs{s}\geq k+k'$, we have $\ccap(\rrg_{k,k'}(s))=0$ if and only if $s$ is periodic with period $\gcd(k,k')$.
\end{theorem}
\begin{IEEEproof}
We start with the easy direction. Assume $s$ is periodic with period $\gcd(k,k')$. Note that in this case $\rrg_{k,k'}(s)$ contains only one sequence of length $ik+k'$ for each $i\ge1$, which is itself a periodic extension of $s$. No other sequences appear in $\rrg_{k,k'}(s)$. Thus, the capacity is $0$.

We now turn to the other direction. Assume the capacity is $0$. We
further assume $s=x_1 \dots x_{k}y_{1}\dots y_{k'}$, with
$x_i,y_i\in\Sigma$, has length $k+k'$. The general case then follows
easily. The proof in this direction is divided into two cases.

For the first case, let $k>k'$, and denote $k''=k-k'$. We show that
$s$ is periodic with period $\gcd(k,k')$.  From Corollary
\ref{cor:hamming}, it follows that $y_{1}\dots y_{k'}=x_1 \dots
x_{k'}$ so we can write $s=x_1 \dots x_{k}x_1 \dots
x_{k'}$. Furthermore, said corollary implies that $x_{i}=x_{k'+i}$ for
$i\in\left[k-k'\right]$ and so $s=x_1 \dots x_{k'}x_1 \dots
x_{k''}x_1 \dots x_{k'}$.  By once applying the rule of
$\rg_{0,k,k'}$ we obtain
\[
t=x_1 \dots x_{k'}\ x_1 \dots x_{k''}\ x_1 \dots x_{k'}\ x_1 \dots x_{k'}\ x_1 \dots x_{k''}.
\]

Now let us apply Corollary \ref{cor:hamming} to the substring
$t'=x_1 \dots x_{k''}\ x_1 \dots x_{k'}\ x_1 \dots x_{k'}$ of
$t$. Since $\ccap(\rrg_{k,k'}(s))=0$, we must have
$\ccap(\rrg_{k,k'}(t))=0$, and obviously, also
$\ccap(\rrg_{k,k'}(t'))=0$. Applying Corollary \ref{cor:hamming} to the
last case of $t'$, we get that
\[
x_1 \dots x_{k''}x_1 \dots x_{k'}=x_1 \dots x_{k'}x_1 \dots x_{k''},
\]
that is, the sequence  $x_1 \dots x_{k''}x_1 \dots x_{k'}$, which has length $k$, equals itself when cyclically shifted by $k'$. Hence, it is periodic with period $\gcd(k,k')$ and thus $s$ is periodic with the same period. 

For the second case, let $k\leq k'$. Denote $x=x_1 \dots x_{k}$ and
$y=y_{1}\dots y_{k'}$, so $s=xy$.  Find integers $q$ and $r$ such that
$k'=qk+r$ and $0\le r<k$ and let $t$ be the sequence obtained from $s$ by $q+1$ times applying $\rg_{0,k,k'}$, that is, 
\begin{align*}
t & =xyx^{q+1}\\
 & =x_{1,k}\,y_{1,k}\,y_{k+1,k}\dots y_{(q-1)k+1,k}\,y_{qk+1,r} \parenv{x_{1,k}}^{q+1}.
\end{align*}
Note that since $\ccap(\rrg_{k,k'}(t))=0$, we also have $\ccap(\rrg_{k,k'}(t'))=0$ for any $(k+k')$-substring $t'$ of $t$. Hence, we can apply Corollary~\ref{cor:hamming} to any $(k+k')$-substring $t'$ of $t$.

For $i=0,1,\dotsc,q-1$, in that order, applying Corollary~\ref{cor:hamming} to the $(k+k')$-substring $t_{ik+1,k+k'}$ implies
that
\begin{equation}
x_{1,k}=y_{ik+1,k}.\label{eq:xy}
\end{equation}
Next, note that from~(\ref{eq:xy}), for the $(k+k')$-substring $t_{qk+1,k+k'}$, we have
\begin{align*}
t_{qk+1,k+k'}&=y_{(q-1)k+1,k}\,y_{qk+1,r}\parenv{x_{1,k}}^q\\
&=x_{1,k}\,y_{qk+1,r}\parenv{x_{1,k}}^q.
\end{align*}
By applying Corollary~\ref{cor:hamming} to this sequence, we find
\[t_{qk+1,k+k'}=x_{1,k}x_{1,r}\parenv{x_{1,k}}^q.\]
Thus, we have 
\[t=(x_{1,k})^{q+1}(x_{1,r})(x_{1,k})^{q+1}.\]
Finally, we apply Corollary~\ref{cor:hamming} to the $(k+k')$-substring 
\[t_{qk+r+1,k+k'}=x_{r+1}\cdots x_k\, x_1\cdots x_r \, x_1\cdots x_k\]
which shows that
\[x_{r+1}\cdots x_k\, x_1\cdots x_r = x_1\cdots x_k.\]
Since $x_1\cdots x_k$ equals itself when cyclically shifted by $r$, it is periodic with period $\gcd(k,r)=\gcd(k,k')$. Hence $t$ is periodic with the same period and so is $s$.

We have shown that for the special case of $|s|=k+k'$, if the capacity is zero, then $s$ is periodic with period $\gcd(k,k')$. Now suppose $|s|>k+k'$ and that $\ccap(\rrg_{k,k'}(s))=0$. Let $d=\gcd(k,k')$ and, for the moment, also suppose that $d$ divides $|s|$. Let \[C=\left\{s_{id+1,k+k'}:0\le i\le \frac{|s|-(k+k')}{d}\right\}\] be a set of $(k+k')$-substrings of $s$ that cover $s$ and each consecutive pair overlap in $d$ positions. Since the capacity for each of these $(k+k')$-substrings is also zero, they are periodic with period $d$. Because of their overlaps and the fact that they cover $s$, it follows that $s$ is also periodic with period $d$. To complete the proof it remains to consider the case in which $d$ does not divide $|s|$. In this case, we can repeat the same argument but with adding the substring $s_{|s|-(k+k')+1,(k+k')}$ to the set $C$ to ensure that $s$ is covered by overlapping $(k+k')$-substrings. 
\end{IEEEproof}

We now turn to discuss the dependence of $\ccap(\rrg_{k,k'}(s))$
on $s$. For a sequence $x\in\Sigma^*$ and two symbols $a,b\in R(x)$, let
\[
\Delta_{x}\parenv{a,b}=\mathset{ j ~\left|~ \exists i, x_{i}=a,x_{i+j}=b\right.}, 
\]
be the set of the differences of positions of $a$ and $b$ in $x$.
Furthermore, let
\[
\rho_{x,\ell}\parenv{a,b}=\mathset{ \parenv{j\bmod\ell}\mid j\in\Delta_{x}\parenv{a,b}} .
\]

\begin{lemma}
\label{lem:capless}
Let $\Sigma$ be some finite alphabet, $d>0$ an integer, and
$D\subset\mathset{0,1,\dots,d-1}$ some subset,
$\abs{D}<d$. Consider the constrained system $S\subseteq \Sigma^*$
such that for every $x\in S$, and every two symbols $a,b\in\Sigma$
(not necessarily distinct), $\rho_{x,d}(a,b) \subseteq D$. Then
$\ccap(S) < \log_2 \abs{\Sigma}$.
\end{lemma}
\begin{IEEEproof}
We begin by constructing a De-Bruijn graph of order $d+1$ over
$\Sigma$, $\cG''(V'',E'')$, defined in the following way. We set
$V''=\Sigma^{d+1}$, and a directed edge connects $v=v_1\dots
v_{d+1}\in V''$ and $v'=v'_1\dots v'_{d+1}\in V''$, if $v'_i=v_{i+1}$
for all $1\leq i\leq d$. That edge has label $v'_{d+1}\in \Sigma$. The
graph is regular with out-degree $\abs{\Sigma}$. Clearly the set of
finite strings read along paths taken in $\cG''$ is simply
$S''=\Sigma^*$. In particular, by Perron-Frobenius theory, if
$A_{\cG''}$ is the adjacency matrix of $\cG''$, since $\cG''$ is
clearly primitive,
\[\ccap(S'')=\log_2 \lambda(A_{\cG''}) = \log_2 \abs{\Sigma}.\]

As the next step, we construct a graph $\cG'(V',E')$ from $\cG''(V'',E'')$ 
by setting $V'=V''$, and removing all edges $v\to u$, such that
\[ \rho_{v,d}(a,b)\cup \rho_{u,d}(a,b) \not\subseteq D,\]
for some $a,b\in\Sigma$. The labels of the surviving edges remain the
same.  We define $S'$ to be the set of strings read from finite paths
in $\cG'$.  Since $\abs{D}<d$, $A_{\cG'}$ is obtained from $A_{\cG''}$
be changing at least one entry from $1$ to $0$. By Perron-Frobenius
theory,
\[\ccap(S') \leq \log_2 \lambda(A_{\cG'}) < \log_2 \lambda(A_{\cG''})
=\log_2 \abs{\Sigma}.\]
Finally, since it is clear that $S\subseteq S'$, we get
\[\ccap(S)\leq \ccap(S') < \log_2 \abs{\Sigma},\]
as claimed.
\end{IEEEproof}

Using Lemma \ref{lem:capless} we obtain the following theorem.
\begin{theorem}
\label{th:capnotmax}
Let $s\in\Sigma^*$ have length at least $k+k'$ and denote $d=\gcd(k,k')$. If, for some $a,b\in R(s)$, we have $\abs{\rho_{s,d}(a,b)}<d$ then $\ccap(\rrg_{k,k'}(s))<\log_2 \delta(s)$.
\end{theorem}
\begin{IEEEproof}
We observe that for any $x,x'\in\rrg_{k,k'}(s)$, and for $a,b\in R(s)$, we have
\[
\rho_{x,d}\parenv{a,b}=\rho_{x',d}\parenv{a,b},
\]
where $d=\gcd(k,k')$. This can be easily seen by noting
that any function in $\Rg_{k,k'}$ changes the differences between
positions of two elements by a linear combination of $k$ and $k'$. 
We then apply Lemma \ref{lem:capless}.
\end{IEEEproof}

\begin{theorem}
For $s\in\Sigma^*$ with $\abs{s}\geq k+k'$, if $\gcd(k,k')=1$, then $\ccap(\rrg_{k,k'}(s))$
depends on $s$ only through $\delta(s)$.
\end{theorem}
\begin{IEEEproof}
The proof is similar to that of Theorem~\ref{thm:mirror3}. In that
light, it suffices to show that in a sequence $y\in\Sigma^*$ of length
$m\ge k+k'$, a symbol $a\in R(y)$ can be ``pushed'' to the end.
That is, we can find a sequence $y''\in\rrg_{k,k'}(y)$ that ends with
$a$.

Suppose $a$ is in position $i$ in $y$. Without loss of generality (similar to Lemma \ref{lem:mirror2.5}), we may assume $i>k$ and $m-i\ge k'-1$. 

Let $y'=\rg_{i-1,k,k'}(y)$. There is a copy of $a$ at position $i'=i$ whose distance from the end of $y'$ is $\abs{y'}-i'=k+m-i$ and this is an increase of size $k$ compared to $y$. We update $y'$ as $y'\leftarrow\rg_{i'-1,k,k'}(y')$. In each step, the distance of $a$ at position $i'$ from the end of $y'$ increases by $k$. We continue until we have $k'\mid\abs{y'}-i'$. This eventually happens as $\gcd(k,k')=1$. 

Now we let $y''=\rg_{i'-k,k,k'}(y')$. There is a copy of $a$ in $y''$ at position $i''=i'+k+k'$. The distance of this copy of $a$ from the end of $y''$ is $\abs{y''}-i''=\abs{y'}-i'-k'$. Thus the distance is decreased by $k'$. We update $y''$ and $i''$ as $y''\leftarrow\rg_{i''-k,k,k'}(y'')$ and $i''\leftarrow i''+k+k'$. We continue until $a$ is the last element of $y''$. The rest of the argument follows along the same lines as those of Lemma~\ref{lem:mirror2.5} and Theorem~\ref{thm:mirror3}.
\end{IEEEproof}

\bibliographystyle{IEEEtranS}
\bibliography{allbib}

\end{document}